\documentclass[a4paper, 12pt]{article}

\usepackage[margin=2cm]{geometry}
\usepackage{hyperref,xcolor}
\usepackage{amsthm, amssymb, amsmath}
\usepackage{enumerate}
\usepackage{algorithmic}
\usepackage{authblk}
\usepackage{graphics}
\newcommand{\Mod}[1]{\ (\mathrm{mod}\ #1)}

\hypersetup{
	colorlinks=false,
	linkcolor=red,
	citecolor=red,
	pdfborder={0 0 0},
}
\date{}

\title{On size of a $r$-wise fractional $L$-intersecting family}

\author[1]{Tapas Kumar Mishra}

\affil[1]
{
	Department of Computer Science and Engineering, \authorcr
	National Institute of Technology, Rourkela 769008, India. \authorcr
	mishrat@nitrkl.ac.in
}

\theoremstyle{definition}
\newtheorem{definition}{Definition}
\newtheorem{example}[definition]{Example}

\theoremstyle{plain}
\newtheorem{theorem}{Theorem}
\newtheorem{lemma}[theorem]{Lemma}

\newtheorem{conjecture}[theorem]{Conjecture}

\theoremstyle{remark}

\newtheoremstyle{plainitshape}
{}
{}
{\itshape}
{}
{\itshape}
{.}
{0.5em}
{}
\theoremstyle{plainitshape}

\newtheoremstyle{cases}
{}
{}
{}
{}
{}
{\newline}
{0.5em}
{{\itshape \thmname{#1}} \thmnumber{#2} ({\itshape\thmnote{#3}}).\medskip}
\theoremstyle{cases}

\newtheoremstyle{constructions}
{}
{}
{}
{}
{}
{}
{0.5em}
{{\itshape \thmname{#1}} \thmnumber{#2}\medskip}
\theoremstyle{constructions}

\definecolor{ao(english)}{rgb}{0.0, 0.5, 0.0}

\newcommand{\rogers}[1]{{\color{black} #1}}

\begin{document}

\maketitle


\begin{abstract}
	Let $L = \{\frac{a_1}{b_1}, \ldots , \frac{a_s}{b_s}\}$, where for every $i \in [s]$, $\frac{a_i}{b_i} \in [0,1)$ is an irreducible fraction. Let $\mathcal{F} = \{A_1, \ldots , A_m\}$ be a family of  subsets of $[n]$. We say $\mathcal{F}$ is a \emph{r-wise fractional $L$-intersecting family} if for every distinct $i_1,i_2, \ldots,i_r \in [m]$, there exists an $\frac{a}{b} \in L$ such that $|A_{i_1} \cap A_{i_2} \cap \ldots \cap A_{i_r}| \in \{ \frac{a}{b}|A_{i_1}|, \frac{a}{b} |A_{i_2}|,\ldots, \frac{a}{b} |A_{i_r}| \}$. In this paper, we introduce and study the notion of r-wise fractional $L$-intersecting families. This is a generalization of  notion of fractional $L$-intersecting families studied in [Niranjan et.al, Fractional $L$-intersecting families, The Electronic Journal of Combinatorics, 2019].
\end{abstract}

\section{Introduction}
\label{sec:Intro}

A family $\mathcal{F}$ of subsets of $[n]=\{1,\ldots,n\}$ is said to be  \emph{$L$-intersecting} if for every $A_i, A_j \in \mathcal{F}$ with $A_i \neq A_j$, we have $|A_i \cap A_j| \in L$. This problem has been studied extensively in literature.
One of the earliest results on the problem is by Ray-Chaudhuri and Wilson \cite{ray1975t}
who proved that  $|\mathcal{F}| \leq {n \choose s}$ provided $\mathcal{F}$ is $t$-uniform. Frankl and Wilson \cite{frankl1981intersection} proved that $|F| \leq {n \choose s} + {n \choose s-1} + \cdots + {n \choose 0}$ when the uniformity restriction on $\mathcal{F}$ is revoked. Alon, Babai and Suzuki \cite{alon1991multilinear} proved the above result using an ingenious linear algebraic argument. In the same paper, the authors generalized the notion of $L$-intersecting families and obtained the following result.
\begin{theorem}\cite{alon1991multilinear}
	Let $L=\{l_1,\ldots,l_s\}$ be a set of $s$ non negetive integers, and $K=\{k_1,\ldots,k_q\}$ be a set of integers satifying $k_i > s-q$ for each $i$. Suppose $\mathcal{A}=\{A_1,\ldots,A_m\}$ be a family of subsets of $[n]$ such that $|A_i| \in K$ for each $1 \leq i \leq m$ and $|A_i \cap A_j| \in L$ for each pair with $i \neq j$. Then,
	\begin{align*}
	m \leq \binom{n}{s}+ \ldots + \binom{n}{s-q+1}.
	\end{align*}
\end{theorem}
This upper bound is tight as given by the family of all subsets of $[n]$ of size between $s-q+1$ and $s$. Gromuluz and Sudakov \cite{grolmusz2002k} extended the results of Frankl-wilson and Alon-Babai-Suzuki to $r$-wise $L$-intersecting families.

\begin{definition}
Let $r\geq 2$ and $L=\{l_1,\ldots,l_s\}$ be a set of $s$ non-negative integers. If $\mathcal{A}=\{A_1,\ldots,A_m\}$ be a family of subsets of $[n]$ such that 
$|A_1 \cap \ldots \cap A_r| \in L$ for every collection of $r$ elements in $\mathcal{A}$, then $\mathcal{A}$ is a $r$-wise $L$-intersecting family.
\end{definition}

\begin{theorem}\cite{grolmusz2002k}
	Let $\mathcal{A}$ be a $r$-wise $L$-intersecting family with $L=\{l_1,\ldots,l_s\}$ where $s\geq 1$ and each $l \in L$ are non-negative integers.
	Then, 
	\begin{align*}
	|\mathcal{A}|\leq (r-1)\left(\binom{n}{s}+ \ldots + \binom{n}{0}\right).
	\end{align*}
	Moreover, if the sizes of every member of $\mathcal{A}$ lies in $K=\{k_1,\ldots,k_q\}$ where each $k_i > s-q$, then
	\begin{align*}
	|\mathcal{A}|\leq (r-1)\left(\binom{n}{s}+ \ldots + \binom{n}{s-q+1} \right).
	\end{align*}
\end{theorem} 

F\H{u}redi and Sudakov \cite{furedi2004} improved the above bound and showed that their bound is asymptotically optimal.

\begin{theorem}\cite{furedi2004}
Let $L$ be a subset of non-negative integers of size $s$, $r \geq 2$ and $\mathcal{A}$ be
a $r$-wise $L$-intersecting family of subsets of an $n$-element set. Then there exists an
integer $n_0=n_0(r,s)$ such that for all $n>n_0$,
\begin{align*}
|\mathcal{A}| \leq \frac{r-s+1}{s+1} \binom{n}{s}+ \sum_{i<s}\binom{n}{i}.
\end{align*}
\end{theorem}

An improvement to the above bound was provided by Kang et.al. \cite{kang2011} who proved the following theorem.

\begin{theorem}\cite{kang2011}
Let $L=\{l_1,\ldots,l_s\}$ be a set of non-negative integers of size $s$ and $\mathcal{A}$ be
a $r$-wise $L$-intersecting family of subsets of an $n$-element set. Then, if $|\cap_{A \in \mathcal{A}} A| < l_1$, $|\mathcal{A}|=o(n^s)$.
Moreover, if if $|\cap_{A \in \mathcal{A}} A| \geq l_1$ and $n$ sufficiently large, 
\begin{align*}
|\mathcal{A}| \leq \frac{r-s+1}{s+1} \binom{n-l_1}{s}+ \sum_{i<s}\binom{n-l_1}{i}.
\end{align*}
\end{theorem}

Various researchers have worked on many variants of the $L$-intersecting families, see \cite{snevily2003sharp,sneville1994,chen2009,liu2009,liu2014set,liu2016,li2016,liu2017,liu2018,wang2018} for detail.

Let $L = \{\frac{a_1}{b_1}, \ldots , \frac{a_s}{b_s}\}$, where for every $i \in [s]$, $\frac{a_i}{b_i} \in [0,1)$ is an irreducible fraction. Let $\mathcal{F} = \{A_1, \ldots , A_m\}$ be a family of  subsets of $[n]$. We say $\mathcal{F}$ is a \emph{fractional $L$-intersecting family} if for every distinct $i,j \in [m]$, there exists an $\frac{a}{b} \in L$ such that $|A_i \cap A_j| \in \{ \frac{a}{b}|A_i|, \frac{a}{b} |A_j|\}$. Niranjan et.al. \cite{niranj2019} introduced the notion of fractional $L$-intersecting families and proved that $m = \mathcal{O}\left(\binom{n}{s} \left( \frac{\log^2n}{\log \log n}\right) \right)$. When $L=\{\frac{a}{b}\}$, the bound on $m$ improves to $\mathcal{O}\left(n \log n\right)$. In this paper, we generalize the notion of fractional $L$-intersecting family to $r$-wise fractional $L$-intersecting family
in the natural way.

\begin{definition}[$r$-wise fractional $L$-intersecting family]
Let $L = \{\frac{a_1}{b_1}, \ldots , \frac{a_s}{b_s}\}$, where for every $i \in [s]$, $\frac{a_i}{b_i} \in [0,1)$ is an irreducible fraction. Let $\mathcal{F} = \{A_1, \ldots , A_m\}$ be a family of  subsets of $[n]$. We say $\mathcal{F}$ is a \emph{$r$-wise fractional $L$-intersecting family} if for every distinct $i_1,i_2, \ldots,i_r \in [m]$, there exists an $\frac{a}{b} \in L$ such that $|A_{i_1} \cap A_{i_2} \cap \ldots \cap A_{i_r}| \in \{ \frac{a}{b}|A_{i_1}|, \frac{a}{b} |A_{i_2}|,\ldots, \frac{a}{b} |A_{i_r}| \}$. 
\end{definition}

In Section \ref{sec:pot2}, we prove the following theorem.

\begin{theorem}
	\label{thm:main1}
	Let $n$ be a positive integer. Let $L = \{\frac{a_1}{b_1}, \ldots,\frac{a_s}{b_s}\}$, where for every $i \in [s]$, $\frac{a_i}{b_i} \in [0,1)$ is an irreducible fraction. Let $\mathcal{F}$ be a $r$-wise fractional $L$-intersecting family of subsets of $[n]$, where $r \geq 3$. 
	Then, $|\mathcal{F}| \leq 2\frac{\ln^2 n}{\ln \ln n}(r-1)\left( \sum_{l=0}^s \binom{n}{l} \right).$ Moreover, the bound improves to $2\frac{\ln^2 n}{\ln \ln n}(r-1)\binom{n}{s}$, if $s \leq n+1 - 2\ln n$.
\end{theorem}

Consider the following examples for a $r$-wise fractional $L$-intersecting family.

\begin{example}
	 Let $L = \{\frac{0}{s}, \frac{1}{s}, \ldots , \frac{s-1}{s}\}$, where $s~(= |L|)$ is a constant. The collection of all the $s$-sized subsets of $[n]$ is a $r$-wise fractional $L$-intersecting family of cardinality ${n \choose s}$. In this case, the bound given by Theorem \ref{thm:main1} is asymptotically tight up to a factor of $(r-1)\frac{\ln^2 n}{\ln\ln n}$.  We believe that if $\mathcal{F}$ is a $r$-wise fractional $L$-intersecting family of maximum cardinality, where $s$ ($=|L|$) is a constant, then $|\mathcal{F}| \in \Theta(rn^s)$.
\end{example}

We note that the linear algebraic techniques which are useful to derive the bounds on
fractional $L$-intersecting families are no longer directly applicable in this case due to the requirements. 
In Section \ref{sec:pot2}, we use a special refinement trick to reduce it into a form such that linear algebraic methods can be used.

Next, we turn our attention to the case when $|L|=s=1$. In the context of classical $L$ intersecting families, when $|L|=s=1$, the Fisher's Inequality (see Theorem 7.5 in \cite{jukna2011extremal}) yields $|\mathcal{F}| \leq n$, where $\mathcal{F}$ is a $L$ intersecting family. Study of such intersecting families was initiated by Ronald Fisher in 1940 (see \cite{fisher1940examination}) in the context of design theory. Analogously, consider the scenario when $L= \{\frac{a}{b}\}$ is a singleton set. Can we get a tighter bound (compared to Theorem \ref{thm:main1})  in this case? We show in Theorem \ref{thm:singleton_b_is_a_prime} that if $b$ is a constant prime we do have a tighter bound. 

\begin{theorem}
	\label{thm:singleton_b_is_a_prime}
	Let $n$ be a positive integer. Let $\mathcal{G}$ be a $r$-wise fractional $L$-intersecting families of subsets of $[n]$, where $L=\{\frac{a}{b}\}$, $\frac{a}{b} \in [0, 1)$, and $b$ is a prime. Then, $|\mathcal{G}| \leq (b-1)(r-1)(n+1)\lceil \frac{\ln n}{\ln b} \rceil + 1$. 
\end{theorem}

Assuming $L= \{ \frac{1}{2}\}$, Examples \ref{examp:bisectionClosedEasy} in Section \ref{sec:singleton} give  $r$-wise fractional $L$-intersecting families on $[n]$ of cardinality $\Omega(n\ln r)$ thereby implying that the bound obtained in Theorem \ref{thm:singleton_b_is_a_prime} is asymptotically tight up to a factor of $r\frac{\ln n}{\ln r}$ when $b$ is a constant prime.    
We believe that the cardinality of such families is at most $crn$, where $c>0$ is a constant.

The rest of the paper is organized in the following way: In Section \ref{sec:pot2}, we give the proof of Theorem \ref{thm:main1} after introducing some necessary lemmas in the beginning.  In Section \ref{sec:singleton}, we consider the case when $L$ is a singleton set and give the proof of Theorem \ref{thm:singleton_b_is_a_prime}.  Finally, we conclude with some remarks, some open questions, and a conjecture.

Before moving on to the proof of Theorem \ref{thm:main1}, we state few key lemmas that will be essential in the proof.

\begin{lemma}[Lemma 13.11 in \cite{jukna2011extremal}, Proposition 2.5 in \cite{babai1992linear}]
	\label{lem:diag_criterion}
	For $i = 1, \dots , m$ let $f_i:\Omega \rightarrow \mathbb{F}$ be functions and $v_i \in \Omega$ elements such that 
	\begin{enumerate}
		\item[(a)]$f_i(v_i) \neq 0$ for all $1 \leq i \leq m$;
		\item [(b)]$f_i(v_j) = 0$ for all $1 \leq j < i \leq m$. 
	\end{enumerate}
	Then $f_1, \ldots , f_m$ are linearly independent members of the space $\mathbb{F}^{\Omega}$. 
\end{lemma}

\begin{lemma}[Lemma 5.38 in \cite{babai1992linear}]
	\label{lem:swallow-2}
	Let $p$ be a prime; $\Omega = \{0,1\}^n$. Let $f \in \mathbb{F}_p^\Omega$ be defined as $f(x) = \sum_{i=1}^n x_i - k$. For any $A \subseteq [n]$, let $V_A \in \{0,1\}^n$ denote its $0$-$1$ incidence vector and let $x_A = \Pi_{j\in A} x_j$. Assume $0\leq s,k \leq p-1$ and $s+k \leq  n$. Then, the set of functions $\{x_Af~:~|A| \leq s-1\}$ is linearly independent in the vector space $\mathbb{F}_p^\Omega$ over $\mathbb{F}_p$. 
\end{lemma}

\section{Proof of Theorem \ref{thm:main1}}
\label{sec:pot2}
Let $\mathcal{F}$ be a $r$-wise fractional $L$-intersecting family of subsets of $[n]$, where $r \geq 3$, $L$ is as defined in the theorem.
Let $p$ be a prime. We partition $\mathcal{F}$ into $p$ parts, namely $\mathcal{F}_0, \ldots , \mathcal{F}_{p-1}$, where $\mathcal{F}_j = \{ A \in \mathcal{F}~:~|A| \equiv j~ (mod~ p)\}$. 
\subsubsection*{Estimating $|\mathcal{F}_j|$, when $j>0$.}

If for every pair of sets $A, B \in \mathcal{F}_j$, $|A \cap B| \in \{\frac{a_1}{b_1}|A|, \ldots, \frac{a_s}{b_s}|A|, \frac{a_1}{b_1}|B|, \ldots, \frac{a_s}{b_s}|B|\}$,
choose the set $A$ with largest cardinality in  $\mathcal{F}_j$, set $X_1=A$ and $Y_1=A$, and remove $A$ from $\mathcal{F}_j$.
Otherwise, there is a collection of $k$ sets $\{A_1,\ldots,A_k\}$ such that $|\cap_{i=1}^k A_i| \not\in \{\frac{a_1}{b_1}|A_1|, \ldots, \frac{a_s}{b_s}|A_1|, \ldots,$  $\frac{a_1}{b_1}|A_k|, \ldots, \frac{a_s}{b_s}|A_k|\}$, and addition of any more set $A$ into $\{A_1,\ldots,A_k\}$ makes $|\cap_{i=1}^k A_i \cap A| \in \{\frac{a_1}{b_1}|A_1|, \ldots, \frac{a_s}{b_s}|A_1|, \ldots, \frac{a_1}{b_1}|A_k|, \ldots, \frac{a_s}{b_s}|A_k|, \frac{a_1}{b_1}|A|, \ldots, \frac{a_s}{b_s}|A|\}$.
Set $X_1=A_1$ and $Y_1=\cap_{i=1}^k A_i$.
Remove $A_1,\ldots,A_k$ from $\mathcal{F}_j$. Repeat the process until no more set is left in $\mathcal{F}_j$. 
Let $X_i,Y_i$ be sets constructed as above, $1 \leq i \leq m$.
Observe  that 
\begin{align}
m \geq \frac{|\mathcal{F}_j|}{r-1}.
\end{align}

Let $X_i = \overline{A}_1$, $Y_i=\{\overline{A}_1,\ldots, \overline{A}_k\}$ for some $k$ and $i$.
By construction, 
\begin{align*}
|X_i \cap Y_i|& = |Y_i|\not\in \{\frac{a_1}{b_1}|\overline{A}_1|, \ldots, \frac{a_s}{b_s}|\overline{A}_1|, \ldots,  \frac{a_1}{b_1}|\overline{A}_k|, \ldots, \frac{a_s}{b_s}|\overline{A}_k|\}, \text{ and } \\
|X_r \cap Y_i| &\in \{\frac{a_1}{b_1}|\overline{A}_1|, \ldots, \frac{a_s}{b_s}|\overline{A}_1|, \ldots, \frac{a_1}{b_1}|\overline{A}_k|, \ldots, \frac{a_s}{b_s}|\overline{A}_k|, \frac{a_1}{b_1}|X_r|, \ldots, \frac{a_s}{b_s}|X_r|\}, \text{ for all $r > i$}.
\end{align*}

With each $X_i$ and $Y_i$, associate the 0-1 incidence vector $x_i$ and $y_i$, where 
$x_i(l)=1$ if and only if $l \in X_i$.
Define $m$ functions $f_1$ to $f_m$, where each $f_j \in \mathbb{F}_p^{\{0,1\}^n}$, in the following way.
\begin{align}
f_i(x)=\left(\left< x,y_i\right> -\frac{a_1}{b_1}j\right) \left(\left< x,y_i\right> -\frac{a_2}{b_2}j\right) \cdots \left(\left< x,y_i\right> -\frac{a_s}{b_s}j\right).
\end{align}
 It follows that 
\begin{align*}
f_i(x_i)=\left(\left< x_i,y_i\right> -\frac{a_1}{b_1}j\right) \left(\left< x_i,y_i\right> -\frac{a_2}{b_2}j\right) \cdots\left(\left< x_i,y_i\right> -\frac{a_s}{b_s}j\right) \neq 0
\end{align*}
for $1 \leq i \leq m$, unless $j=0$.
Moreover, $f_i(x_r)=0$ for $1 \leq i< r\leq  m$.
Using Lemma \ref{lem:diag_criterion}, it follows that the multilinear polynomials $f_1,\ldots,f_m$ are linearly independent over $\mathbb{F}_p^{\{0,1\}^n}$. The dimension of the space is $\sum_{l=0}^s \binom{n}{l}$. Therefore, $\sum_{l=0}^s \binom{n}{l} \geq m \geq \frac{|\mathcal{F}_j|}{r-1}$. This implies that
$|\mathcal{F}_j| \leq (r-1)\left( \sum_{l=0}^s \binom{n}{l} \right)$.

The maximum value of $j$ is $p-1$ and we will show shortly that the maximum value of $p$ needed in the proof is $2\ln n$. So, choosing $s \leq n+1 - 2\ln n$, the requirements of Lemma \ref{lem:swallow-2} are satisfied.
We can now improve the upper bound on $|\mathcal{F}_j|$ by using the swallowing trick and Lemma \ref{lem:swallow-2} to prove that $\{f_i~:~1 \leq i \leq m\} \cup \{x_Af~:~ |A| < s \}$ (where $f(x) = \sum_{i=1}^n x_i - j$) is a collection of functions that is linearly independent in the vector space  $\mathbb{F}_p^{\{0,1\}^n}$ over $\mathbb{F}_p$. These functions can be obtained as a linear combination of distinct monomials of  degree at most $s$. This implies that $\sum_{l=0}^s \binom{n}{l} \geq m+\sum_{l=0}^{s-1} \binom{n}{l}$, that is $m \leq \binom{n}{s}$. This yields  
$|\mathcal{F}_j| \leq (r-1) \binom{n}{s} $.

From the discussion above, it is clear that
\begin{align}
|\mathcal{F}_j| \leq \begin{cases}
(r-1) \binom{n}{s}, \text{ if }s \leq n+1 - 2\ln n \\
(r-1)\left( \sum_{l=0}^s \binom{n}{l} \right), \text{ otherwise}
\end{cases}
\text{ for $j > 0$.}
\end{align}

\subsubsection*{Estimating $|\mathcal{F}_0|$.}

In order to estimate $|\mathcal{F}_0|$, we choose a collection $p_1 < p_2 < \ldots <p_t$ of $t$ smallest primes such that $p_1p_2 \ldots p_t >n$. This implies that every set $F$ in $\mathcal{F}$ has a prime $p$ such that $p \nmid |F|$ - that is, $F$ will be counted in the estimation of $|\mathcal{F}_1 \cup \ldots \cup \mathcal{F}_{p-1}|$. 
So, 
\begin{align}\label{ineq:2}
|\mathcal{F}| \leq 
\begin{cases}
t*(p_t-1)(r-1)\binom{n}{s}, \text{ if }s \leq n+1 - 2\ln n, \\
t*(p_t-1)(r-1)\left( \sum_{l=0}^s \binom{n}{l} \right), \text{ otherwise.}
\end{cases}
\end{align}

Now, the only thing that remains is to estimate $t$ and $p_t$.
The product of the first $t$ primes is the \emph{primorial function} $p_t\#$ and it is known that 
$p_t\#= e^{(1+o(1))t\ln t}$. 
Setting $p_t\#= e^{(1+o(1))t\ln t} > n$, we get
$t \leq \frac{\ln n}{\ln \ln n}$. Moreover, using the Prime Number Theorem (see Section 5.1 of \cite{victor2009}),
the $t$th largest prime is at most $2t\ln t$.
Using these facts and Inequality \ref{ineq:2}, Theorem \ref{thm:main1} follows.
\qed


\section{$L$ is a singleton set}
\label{sec:singleton}

As explained in Section \ref{sec:Intro}, Fisher's Inequality is a special case of the classical $L$-intersecting families, where $|L| = 1$. In this section, we study $r$-wise fractional $L$-intersecting families with $|L| = 1$; a fractional variant of the Fisher's inequality.
\subsection{Proof of Theorem \ref{thm:singleton_b_is_a_prime}}
\emph{Statement of Theorem \ref{thm:singleton_b_is_a_prime}:} Let $n$ be a positive integer. Let $\mathcal{G}$ be a $r$-wise fractional $L$-intersecting families of subsets of $[n]$, where $L=\{\frac{a}{b}\}$, $\frac{a}{b} \in [0, 1)$, and $b$ is a prime. Then, $|\mathcal{G}| \leq (b-1)(r-1)(n+1)\lceil \frac{\ln n}{\ln b} \rceil + 1$.

\begin{proof}
	It is easy to see that if $a=0$, then $|\mathcal{G}| \leq n$ with the set of all singleton subsets of $[n]$ forming a tight example to this bound. So assume $a \neq 0$. Let $\mathcal{F} = \mathcal{G}\setminus \mathcal{H}$, where $\mathcal{H} = \{A \in \mathcal{G}~:~b \nmid |A|\}$. From the definition of a $r$-wise fractional $\frac{a}{b}$-intersecting family it is clear that $|\mathcal{H}|\leq 1$. The rest of the proof is to show that $|\mathcal{F}| \leq (b-1)(r-1)(n+1)\lceil \frac{\ln n}{\ln b} \rceil$. We do this by partitioning $\mathcal{F}$ into $(b-1)\lceil \log_b n \rceil$ parts and then showing that each part is of size at most $(r-1)(n+1)$.
	 We define $F_i^j$ as 
	\[\mathcal{F}_i^j= \{A \in \mathcal{F}| |A| \equiv j \Mod i\}.\]
	Since $b$ divides $|A|$, for every $A \in \mathcal{F}$, under this definition $\mathcal{F}$ can be partitioned into families $\mathcal{F}_{b^k}^{ib^{k-1}}$, where $2 \leq k \leq \lceil \log_b n \rceil$ and $1 \leq i \leq b-1$. We show that, for every $i \in [b-1]$ and for every $2 \leq k \leq \lceil \log_b n\rceil$, $|\mathcal{F}_{b^k}^{ib^{k-1}}| \leq (r-1)(n+1)$. 

In order to estimate $|\mathcal{F}_{b^k}^{ib^{k-1}}|$, for each $A \in \mathcal{F}_{b^k}^{ib^{k-1}}$, create a vector $X_A$ as follows:

\[X_A(j)=\begin{cases}
\frac{1}{\sqrt{b^{k-2}}} \text{, if $j \in A$;}\\
0 \text{, otherwise.}
\end{cases} \]

\begin{definition}
Let $x^1,\ldots,x^r \in \mathbb{F}^n$ for some field $\mathbb{F}$, where $x^i=(x^i_1,\ldots, x^i_n)$. The $r$-wise dot product, denoted as $\left \langle x^1,\ldots,x^r\right \rangle$ is defined as
$
\left \langle x^1,\ldots,x^r\right \rangle = \sum_{i=1}^n x^1_i x^2_i \ldots x^r_i.
$
\end{definition}

\rogers{Note that, for distinct sets $A_1,\ldots, A_r \in \mathcal{F}_{b^k}^{ib^{k-1}}$}
\begin{align}
\label{eq:do}
\left \langle X_{A_j},X_{A_j}\right \rangle \equiv b \Mod {b^2},\nonumber \\
\left \langle X_{A_1},\ldots, X_{A_r}\right \rangle \equiv ai \Mod b.
\end{align}

\subsubsection*{Estimating $|\mathcal{F}_{b^k}^{ib^{k-1}}|$}

If for every pair of sets $A, B \in \mathcal{F}_{b^k}^{ib^{k-1}}$, $|A \cap B| \equiv ai \Mod b$,
choose the set $A$ with largest cardinality in  $\mathcal{F}_{b^k}^{ib^{k-1}}$, set $C_1=A$ and $D_1=A$, and remove $A$ from $\mathcal{F}_{b^k}^{ib^{k-1}}$.
Otherwise, there is a collection of $k$ sets $\{A_1,\ldots,A_k\}$ such that $|\cap_{j=1}^k A_j| \not\equiv ai \Mod b$, and addition of any more set $A$ into $\{A_1,\ldots,A_k\}$ makes $|\cap_{j=1}^k A_j \cap A| \equiv ai \Mod b$.
Set $C_1=A_1$ and $D_1=\cap_{j=1}^k A_j$.
Remove $A_1,\ldots,A_k$ from $\mathcal{F}_{b^k}^{ib^{k-1}}$. Repeat the process until no more set is left in $\mathcal{F}_{b^k}^{ib^{k-1}}$. 
Let $C_j,D_j$ be sets constructed as above, $1 \leq j \leq m$.
Observe  that 
\begin{align}
\label{eq:primary}
m \geq \frac{|\mathcal{F}_{b^k}^{ib^{k-1}}|}{r-1}.
\end{align}

Let $C_j = \overline{A}_1$, $D_j=\{\overline{A}_1,\ldots, \overline{A}_k\}$ for some $k$ and $j$.
By construction, $|C_j \cap D_j| = |D_j|\not\in \{\frac{a}{b}|\overline{A}_1|, \ldots, \frac{a}{b}|\overline{A}_k|\}$, and $|C_r \cap D_j| \in \{\frac{a}{b}|\overline{A}_1|, \ldots, \frac{a}{b}|\overline{A}_k|,\frac{a}{b}|C_r|\}, $ for all $r > j$.
From the definition of $\mathcal{F}_{b^k}^{ib^{k-1}}$, Equation \ref{eq:do}, and construction above, it follows that
for any $1 \leq j,l \leq m$,
\begin{align}
\left \langle X_{C_j},X_{D_l}\right \rangle \begin{cases}
\not\equiv ai \Mod {b}, \text{ if $j=l$, } \nonumber\\
\equiv ai \Mod b, \text{ if $j>l$, }
\end{cases}
\end{align} 

Define $m$ functions $f_1$ to $f_m$, where each $f_j \in \mathbb{F}_b^{\mathbb{R}^n}$, in the following way.

$$ f_j(x) = (\langle x, X_{D_j}\rangle - ai). $$

It follows that
\begin{align*}
f_j(X_{C_r})\begin{cases}
\neq 0, \text{ if $j=r$, } \nonumber\\
=0, \text{ if $r>j$, }
\end{cases}
\end{align*}

So, $f_j$'s are linearly independent in the vector space ${\mathbb{F}_b}^{\mathbb{R}^n}$ over ${\mathbb{F}_b}$
(by Lemma \ref{lem:diag_criterion}). Each $f_j$ is thus an appropriate linear combination of distinct monomials of degree at most one.
Therefore, $m \leq \sum_{j=0}^1 {n \choose j}=n+1$. Thus, using Equation \ref{eq:primary},
$|\mathcal{F}_{b^k}^{ib^{k-1}}| \leq (r-1)(n+1)$.
This concludes the proof of the Theorem. 

\end{proof}

We shall call $\mathcal{F}$ a \emph{$r$-wise bisection closed family} if $\mathcal{F}$ is a fractional $L$-intersecting family where $L=\{\frac{1}{2}\}$. We have the following construction that yields a $r$-wise bisection closed family of cardinality at least  $n\{1+\frac{1}{2}+\ldots+ \frac{1}{r}\}-2r$ on $[n]$. 
\begin{example} 
	\label{examp:bisectionClosedEasy}
Let $n$ be an even positive integer. 
		Let $\mathcal{B}_1$ denote the collection of 2-sized sets that contain only 1 as a common element in any two sets, i.e. $\{1,2\}, \{1,3\}, \ldots, \{1,n\}$; and
	let  $\mathcal{B}_2$ denote collection of 4-sized sets that contain only $\{1,2\}$ as common elements, i.e. $\{1,2,3,4\},$ $ \{1,2,5,6\}, \ldots, \{1,2,n-1,n\}$.
Similarly, let $\mathcal{B}_i$ denote collection of $2i$-sized sets that contain only $\{1,2,\ldots,i\}$ as common elements, i.e. $\{1,2,\ldots,i,i+1,\ldots,2i\},$ $ \{1,2,\ldots,i,2i+1,\ldots,3i\}, \ldots, \{1,2,\ldots,i,n-i+1,\ldots,n\}$, for $1 \leq i \leq r$ (possibly excluding the last set in the family if it is not of size $2i$).
	It is not hard to see that $\mathcal{B}_1 \cup \mathcal{B}_2 \cup \ldots \cup \mathcal{B}_r$ is indeed $r$-wise bisection closed.
\end{example}

\section{Discussion}

In this paper, we introduce and study the notion of $r$-wise fractional $L$-intersecting families, which is a generalization of  notion of fractional $L$-intersecting families studied in \cite{niranj2019}. 
If $L=\{\frac{a_1}{b_1}, \ldots,\frac{a_s}{b_s}\}$, Theorem \ref{thm:main1} gives an upper bound of 
$ \mathcal{O}\left( \frac{\ln^2 n}{\ln \ln n}r\binom{n}{s} \right)$ on the size of such families.
When $L$ is a singleton set, this translates to an upper bound of $\mathcal{O}\left(  r n  \frac{\ln^2 n}{\ln \ln n}  \right)$ on the size of such families. 
If $L=\{\frac{a}{b}\}$, $\frac{a}{b} \in [0, 1)$, and $b$ is a prime, Theorem \ref{thm:singleton_b_is_a_prime} gives an upper bound of $\mathcal{O}\left(rn{\ln n}\right)$
We believe that in this case, the upper bound should be linear which we pose as an open problem.
\begin{conjecture}
	Let $\mathcal{F}$ be an $r$-wise fractional $L$-intersecting family, where $L=\{a/b\}$.
	Then, $|\mathcal{F}| = \mathcal{O}\left(rn\right)$.
\end{conjecture}
Let $r$ be a fixed constant and $L = \{\frac{0}{s}, \frac{1}{s}, \ldots , \frac{s-1}{s}\}$, where $s$ is a constant. The collection of all the $s$-sized subsets of $[n]$ is a $r$-wise fractional $L$-intersecting family of cardinality ${n \choose s}$. In this case, the bound given by Theorem \ref{thm:main1} is asymptotically tight up to a factor of $\frac{\ln^2 n}{\ln\ln n}$. We believe that in this case, $|\mathcal{F}| \in \Theta(n^s)$ and improving the bound in Theorem \ref{thm:main1} remains open.

In Theorem 6 and Theorem 8 of \cite{niranj2019}, the authors have shown linear upper bound for fractional $L$-intersecting families for large sized sets and sets of size nearly $\frac{n}{2}$, respectively. Obtaining similar bounds in the case of $r$-wise fractional $L$-intersecting families remains open.

\small
\bibliographystyle{plain}

\end{document}